# Assessing Simulations of Imperial Dynamics and Conflict in the Ancient World

Jim Madge, Giovanni Colavizza, James Hetherington, Weisi Guo, Alan Wilson.

## Abstract

The development of models to capture large-scale dynamics in human history is one of the core contributions of cliodynamics. Most often, these models are assessed by their predictive capability on some macro-scale and aggregated measure and compared to manually curated historical data. In this report, we consider the model from Turchin et al. (2013), where the evaluation is done on the prediction of "imperial density": the relative frequency with which a geographical area belonged to large-scale polities over a certain time window. We implement the model and release both code and data for reproducibility. We then assess its behaviour against three historical data sets: the relative size of simulated polities vs historical ones; the spatial correlation of simulated imperial density with historical population density; the spatial correlation of simulated conflict vs historical conflict. At the global level, we show good agreement with population density ($R^2<0.75$), and some agreement with historical conflict in Europe ($R^2<0.42$). The model instead fails to reproduce the historical shape of individual polities. Finally, we tweak the model to behave greedily by having polities preferentially attacking weaker neighbours. Results significantly degrade, suggesting that random attacks are a key trait of the original model. We conclude by proposing a way forward by matching the probabilistic imperial strength from simulations to inferred networked communities from real settlement data.

## Introduction

Recent work has highlighted how complex societies all share, to some extent, a set of evolutionary characteristics (Turchin et al. 2018). Consequently, the causal factors which (partially) explain how they came to be might be (partially) shared in turn. In particular, competition (e.g., war) and cooperation (e.g., trade) across polities play a crucial role in the emergence, evolution or disappearance of complex societies. Previous work has considered simple and relatively isolated systems of coupled equations in order to model population and warfare dynamics (Turchin et al. 2006; Wilson 2016). Agent-based models have also been considered in order to simulate the emergence of cooperative social behavior (Burtsev and Turchin 2006) while having found wider application in the social sciences (Fonoberova et al. 2012; Gavrilets and Fortunato 2014).

We start with a brief summary of Turchin et al. (2013), which in turn consolidated and developed on substantial previous work (Turchin et al. 2006; Turchin and Gavrilets 2009; Turchin 2009b; Turchin 2011; Turchin 2013), including a simpler agent-based model with similar objectives (Gavrilets et al. 2010). The central premise of Turchin et al. (2013)'s work is to put forward and test a model for the evolution of large-scale complex societies in the Old World (1500 BCE to 1500 CE, approximately). Their key hypothesis is that intense competition in the form of warfare led to the development of institutions which supported large-scale complex societies. It is a given that societies compete, warfare being just one way. Warfare intensity is in turn hypothesized to depend primarily on the spread of military technologies and on geographic factors, such as terrain. Other reasonable and competing hypotheses have been put forward (Gowdy and Krall 2016; Norenzayan et al. 2016; Richerson et al. 2016). For a recent review, see Mesoudi (2017). Turchin et al. (2013) consider cultural multilevel selection as their theoretical framework: there is a natural tendency for complex societies to lose their ultrasocial institutions to the advantage of local ones, in the absence of external competition demanding to muster the resources of the given society more broadly. "Ultrasocial" here stands for any kind of coordination effort, e.g., institutions, happening among not genetically proximally correlated individuals (i.e. beyond the family, tribe, etc.).

In the proposed model, the Afro Eurasian landmass is divided into 100x100km squares, each characterized by its biome (desert, steppe or agriculture) and average elevation. The range of agriculture cells expands over time, according to historical processes. Each cell possesses an individual polity at the beginning of the simulation, while military technologies are endowed to cells adjacent to the steppes and then gradually and diffuse out via conflict. Military technologies are represented as a boolean vector in every cell. Each cell is further endowed with an ultrasocial traits boolean vector, which can be gained by occupying cells with existing traits or with a very low probability over time, and which naturally disappear (with higher probability than appearance) over time, according to the cultural multilevel selection framework. Polities can wage war, giving a chance to enlarge their pool of controlled cells and to acquire or spread ultrasocial traits. War happens randomly over borders of agricultural to agricultural cells and is decided by the relative size of the two polities and their ultrasociality traits. Military technologies, which diffuse over time from the steppes to agricultural cells, increase the probability for a newly conquered cell to undergo ethnocide, which causes the ultrasociality traits of the winner to be copied in the newly conquered cell. The causal chain embedded in such a model is as follows: spread of military technologies, intensification of warfare, evolution of ultrasocial traits, emergence of large-scale societies.

The model was tested empirically by its capacity to simulate how frequently cells belonged to large-scale polities over time, a.k.a the "imperial density" map, using the coefficient of determination ($R^2$) of a linear regression between the true and simulated data for every cell. A large-scale polity is considered as one with a controlled territory of at least 100.000km$^2$ (10 cells) for at least 100 contiguous years. As a consequence of expansionist warfare, cells part of large polities are also more likely to possess ultrasocial traits. Turchin et al. (2013)'s full model accounts for 65% of the variance in historical data, while a model without elevation (a key geographic feature) explains 48% of it. A model where military technologies are seeded randomly or with equal intensity everywhere is not predictive, as is a model where military technology has no impact on ethnocide. These last tests are particularly informative, as they rule out the possibility of artefact results caused by the shape of the grid and further reinforce the authors' hypothesis that steppes and military technology diffusion are key drivers. In Turchin et al. (2013)'s simulation, time is discrete and each time-step is of 2 years (3000 years in total). The full choice of parameters was found by using a mix of relatively uninformative priors and grid-search fitting against ground truth. Importantly, polities can collapse with low probability, positively correlated with their size and negatively correlated with their ultrasocial traits. Further tests controlling for spatial autocorrelation highlight how the single best predictive variables are horse warfare and distance from the steppes, supporting the initial hypothesis of the authors. While Turchin at al. (2013)'s model has good explanatory power with respect to average imperial density, it remains unknown whether it is able to account for the following historical aspects: size and shape of large-scale polities, relation of large-scale polities with densely inhabited areas (cities) and actual areas of conflict.

In this report, our goal is threefold: a) to implement and reproduce Turchin et al. (2013)'s model and results, releasing code and data openly; b) to assess its capacity to simulate further important historical aspects rather than just average imperial density; c) to suggest possible future directions for improvement in view of this assessment. Our intent here is thus to open and problematize, suggesting possible improvements.

We note that subsequent work from Bennett (2016), expanding from Bennett (2015), proposed a model which explicitly included demographic-structural factors, in order to address the following limitations in Turchin at al. (2013): demographic pressure and its relation to structural aspects of states and internal collapse due to civil or inter-state war. While this model is able to simulate well the total area and population part of large-scale polities, it does not embed any change which would improve results on the above-mentioned aspects. As a consequence, we focus on assessing Turchin et al. (2013)'s model here.

Our contribution is organized as follows. We start by assessing our replication of results from Turchin et al. (2013), then expand on the assessment of their model. We further explore results for a greedy version of the model, and conclude by suggesting directions for future work. More details on our implementation of Turchin et al. (2013)'s model can be found in the appendix.

## Baseline evaluation

We first verified our implementation by comparing imperial density results, averaged over 20 independent simulations in the date ranges 1500 BC to 500 BC, 500 BC to 500 AD and 500 AD to 1500 AD, with those of Turchin et al. (2013)'s.

**Figure 1** shows the spatial distribution of imperial density for each of the aforementioned three eras. The left column is the historical imperial density, the centre is Turchin et al. (2013)'s results and the right ours. From visual inspection of the two rightmost columns, it is clear that our simulation very closely matches the original imperial density profile. Across all eras the locations and shapes of regions of high imperial density are consistent. Furthermore the spread of imperial density outwards from the steppes regions in our simulations occurs at the same pace as the original model. Even more subtle patterns such as the distribution of imperial density in sub-Saharan Africa and its lack on the Tibetan plateau and Himalayas are well reproduced.

To further validate our simulations we carried out a linear regression analysis of the historical imperial density for each era against our simulated imperial density for the same era. In the regression, all land cells were considered (*i.e.* including desert cells which are forbidden from belonging to polities). **Table 1** shows the $R^2$ values for each regression alongside those obtained by Turchin et al. (2013). Our values compare favourably to those reported in the original work with the exception of the era 1500 BC to 500 BC where our correlation is higher.

| era | $R^2$ Turchin | $R^2$ this work | $R^2$ greedy attack model |
| --- | --- | --- | --- |
| 1500 BC – 500 BC | 0.56 | 0.66 | 0.22 |
| 500 BC – 500 AD | 0.65 | 0.66 | 0.39 |
| 500 AD – 1500 AD | 0.48 | 0.49 | 0.26 |

**Table 1. Correlation value with real imperial strength data for Turchin and our work.**

Despite not perfect, we consider our replication of Turchin et al. (2013)'s results sufficient for our purposes. Our justification for this is twofold. First, we consider the qualitative comparison of the model (**Figure 1**) to be very strong and our implementation therefore accurately models all important aspects of the original model. Second, the model has a large number of parameters which might justify minor differences in results. These, combined with some discrepancies in the description of the model in the paper and in the original code still present a number of variations to investigate when seeking better fits (see Appendix).

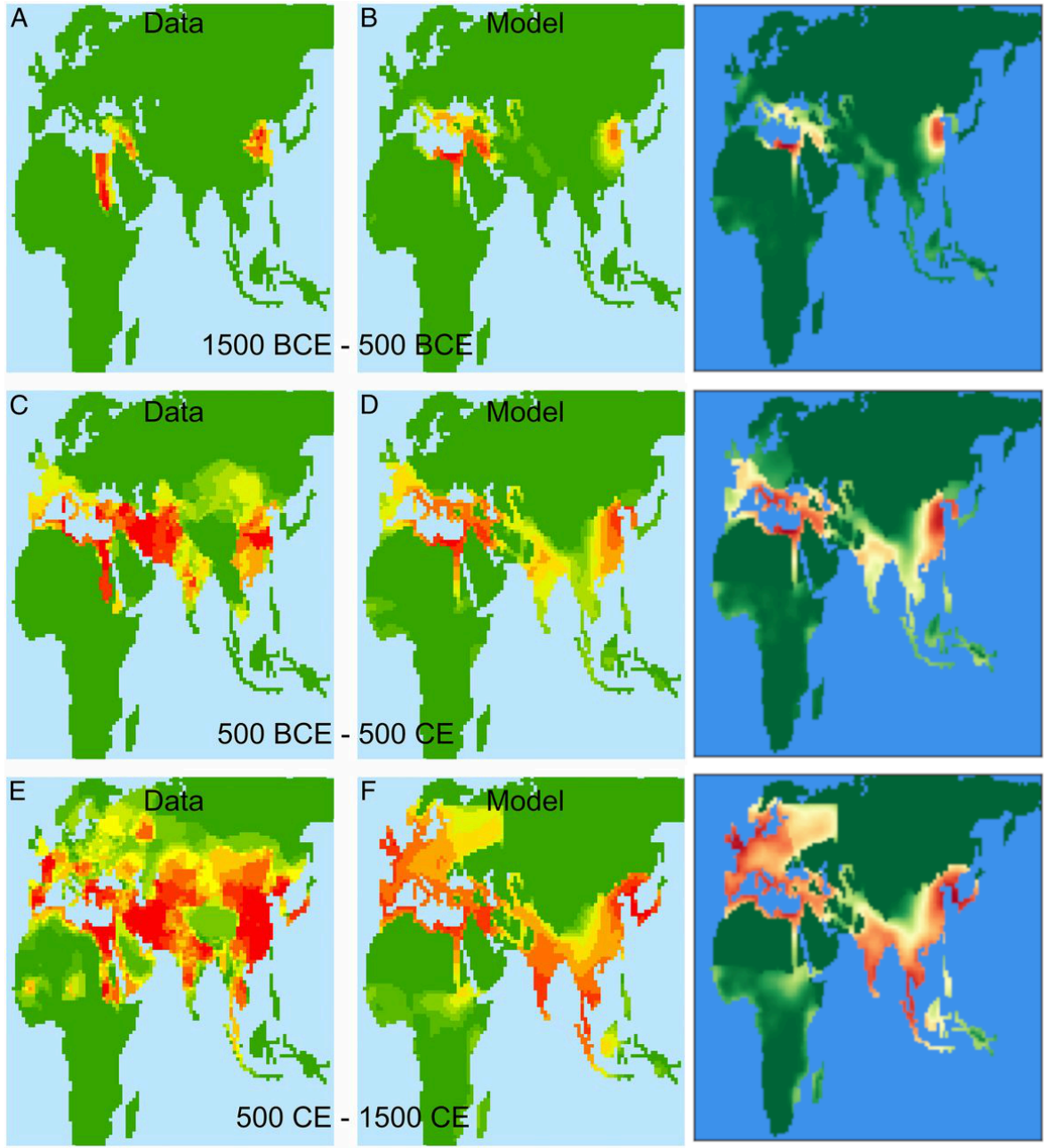

**Figure 1. Imperial density data (left), Turchin et al. (2013) simulated model (centre), our reproduction of the simulated model (right); for three time epochs.**

## Extended evaluation

After having confirmed that our implementation closely replicated the results of the original model, we moved on to further validate the model using historical data sets. Firstly, we sought to investigate the shape and distribution of polities compared to known historical empires. Initially, we surveyed the polities within the historical extent of the Roman empire, as determined by annotated data from Turchin et al. (2013). This region was determined from the same historical polities data used to determine the historical imperial density in

Turchin et al. (2013). From our initial simulations it became clear that in any given simulation it is unreasonable to expect the polities to closely match historical empires in size, shape or lifetime (also cf. Bennett 2016). In fact, the simulation tends to be dominated by a large number of small or single-cell polities at any time. Two typical histograms of the distribution of polity sizes within the historical Roman empire in the years 100 AD and 300 AD are given in **Figure 2**. The left column gives the number of polities of a given size, whereas the right gives the number of cells in polities of a given size. From these plots it is clear that large polities are quite rare and though they may grow to a size where they contain a large number of the available cells they are still typically dwarfed by small and single-cell polities. This effect is driven by the high probability of polity disintegration, at least 5% at every step.

**Figure 2. Polity frequency distribution (left), polity size distribution (right); for different time periods of the area in the Roman empire.**

As a result of this observation, we decided not to continue pursuing analysis involving the specific shape or size of individual polities. However, we do feel that this process was useful in outlining the limitations and appropriate application of the original model. While we cannot expect specific polities in a single simulation to represent any particular historical state or empire, it is clear that the collective distribution of large polities over repeated simulations, expressed as the imperial density, closely matches the historical evolution of large-scale polities. It is therefore important to understand the statistical nature of the model when seeking to validate or make predictions from simulations. Nonetheless, we do discuss in the paper's final section that the distribution of polity sizes potentially represents the vulnerability of a larger polity to fractionalize and therefore that a probabilistic measure of polity fractionalization is needed.

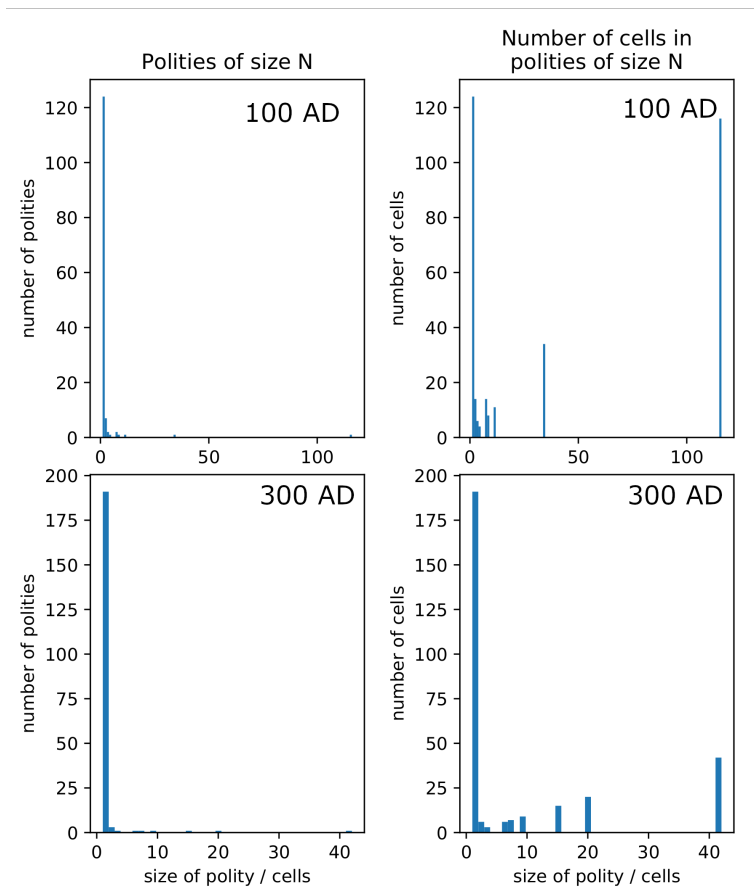

Next, we attempted to further validate the model by comparing imperial density to historical population records from Reba et al. (2016). The dataset contained population records or estimates for cities spanning from 2500 BC to 1975 AD in intervals of uneven size (for example 2500 BC–1000 BC, 0–500 AD and 1400 AD–1500 AD). Where the population data is missing for certain time periods, we have used linear regression to fill the gaps in. Each record in the data set also contained the location of the city as a longitude and latitude. Using these coordinates, each city was projected onto the world map. If two or more cities were found to be within the same cell on the map, their populations were summed. This resulted in a sparse population map with a large number of cells having no population at all. In order to better match population data to the simulated imperial density, a Gaussian blur on the population per cell was applied (with a standard deviation of three cells). We also decided to compare the population data against cumulative imperial density. This reflects that areas of high population do not simply occur immediately when a large state forms but rather grow more quickly in stable regions than those in constant flux.

The blurred population density maps and correlation plots of population with cumulative imperial density are shown in **Figure 3**, for the periods 0–500 AD, 1000 AD–1100 AD and 1400 AD–1500 AD (time ranges as permitted by the population data). For each correlation plot, the $R^2$ value is given on the top edge. There is a strong correlation between cumulative imperial density and population. For all time periods considered $R^2$ values were greater than 0.5 and increasing over time to a maximum of 0.75 in the period 1400 AD–1500 AD. This may reflect an increasing quality of population data in more recent times. The data points with zero imperial density but non-zero populations in **Figure 3** mostly correspond to cells defined as deserts in the model. These cells are currently forbidden from belonging to a polity.

Finally, we considered whether the model could be used to identify regions of conflict. Our first point of investigation was to count how many times each cell is attacked over a simulation. The attack events were compared to a data set of historical battles created from a concatenation of two data sources, both derived from Wikipedia.[1] As with the population data, the battles were projected onto the old world map and blurred. The correlation of attack events with historical battles is in general poor. This is mainly due to a strong relationship between imperial density and the frequency of attacks. Areas which tend to feature large polities see fewer attacks as cells are forbidden from attacking each other if they belong to the same polity. Conversely, areas where large polities are rare see a large number of attack events. In particular sub-Saharan Africa displayed a very large number of attack events in contrast to a small number of documented historical battles. As with population data, it must be noted that the battle dataset might be limited and have better coverage for Western historical battles.

A significantly better correlation was achieved, in this case, by restricting the cells being compared to those located in Europe, the Middle East and North Africa, which is the predominant theatre for the western classical world and medieval Europe and the Levant. On the world stage, this kernel largely eliminates regions with low imperial density and high numbers of attacks, and likely focuses on an area for which battle data is better overall. The left column of **Figure 4** shows the blurred historical battle frequency, the centre column displays the frequency of simulated attack events and the right column shows the correlation between these two on a log-log scale. While better than the correlation for the entire old world, the quality of these fits is below those for population data, with $R^2$ values typically ranging between 0.1 and 0.4. Similarly to the correlations in **Figure 3**, we observe that $R^2$ tends to increase over time, likely reflecting an increase in the quality and number of data points available. Since battles occur at the edges of polities, we speculate that, over a number of simulations the frequency of attacks may convey an intuition of the likely boundaries of empires where land is contested and conflict is more likely. It appears that, as we discovered earlier, the high variance in polity size and shape between simulations may limit how accurate this result can be.

---

[1] https://nodegoat.net/blog.p/82.m/14/a-wikidatadbpedia-geography-of-violence.

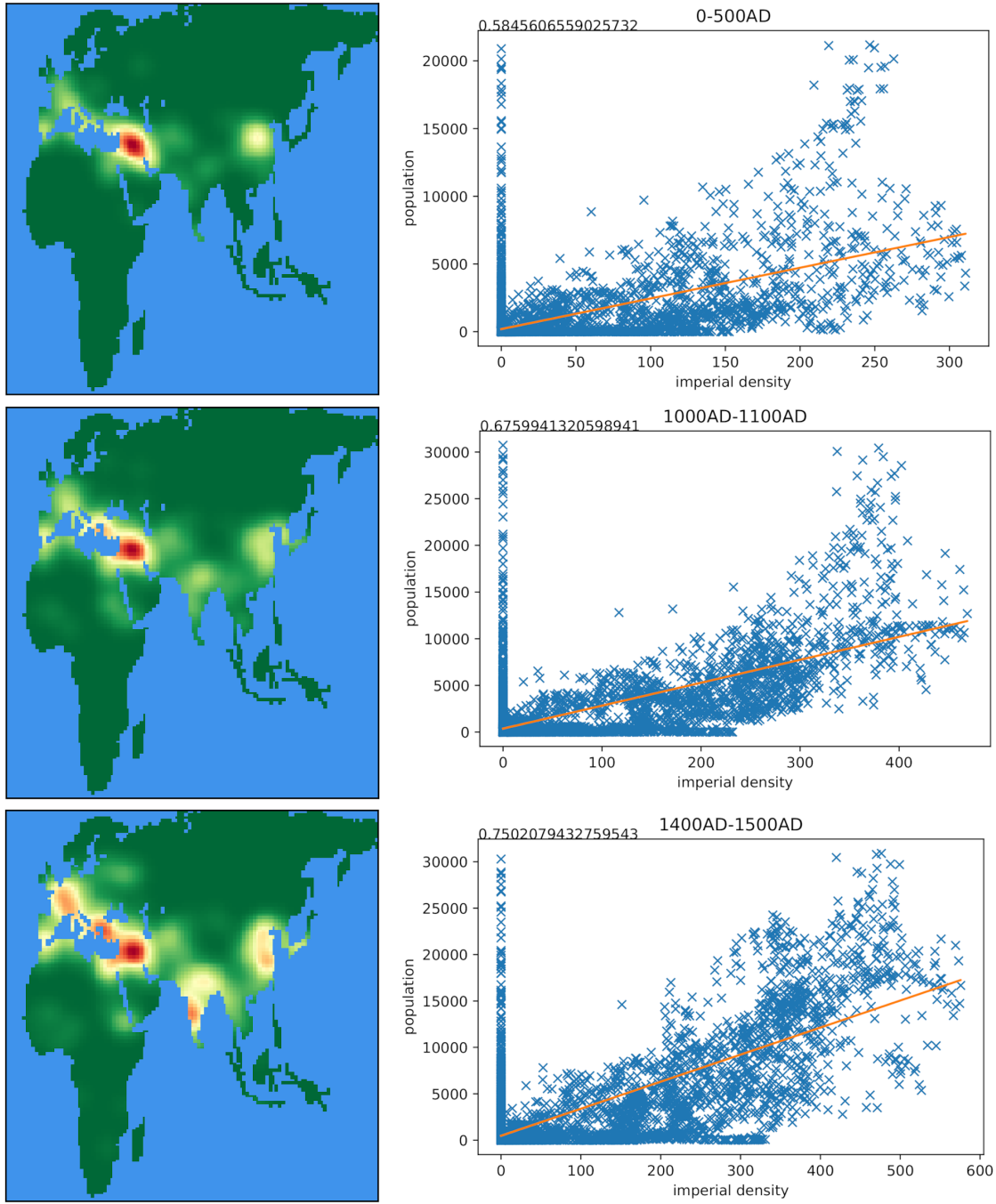

**Figure 3. Historical population data, with Gaussian blur (left), their correlation with cumulative imperial density with a linear fit (right); for different time epochs.**

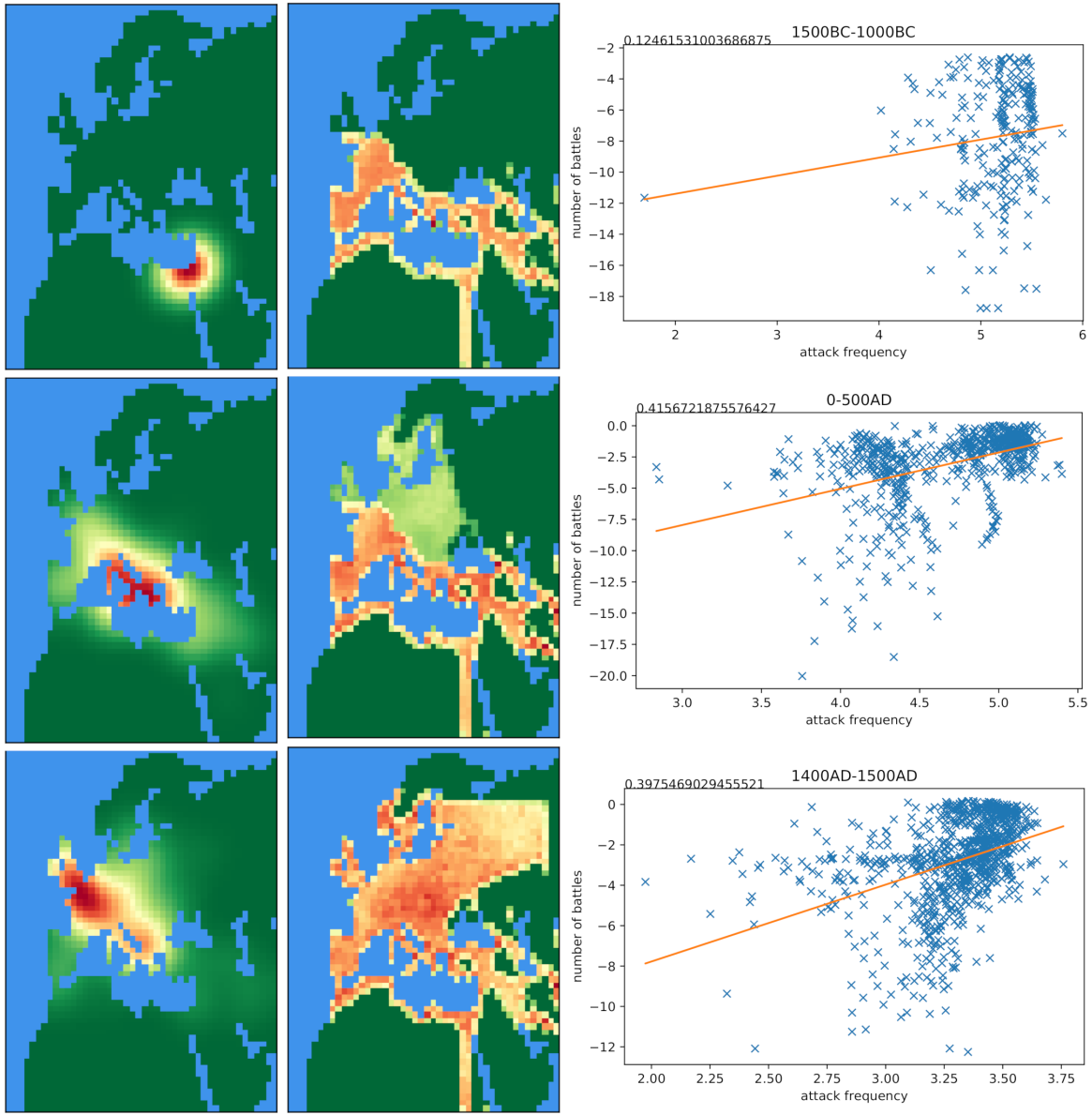

**Figure 4. Historical battles data, with Gaussian blur (left), simulated attack frequency (centre), their correlation with a linear fit, on a log-log scale (right); for different time epochs.**

## Greedy model

The model was extended with the introduction of a 'greedy' method for the selection of an attack target. Using this method, polities are more likely to attack their weaker neighbours. Polities therefore choose their targets more intelligently; they are less likely to engage in battles where victory is unlikely and are more likely to exploit weak neighbours. The greedy model proceeds as follows. First, all neighbours of the attacking polity are enumerated, both land and sea neighbours. Next, the advantage of each neighbour is calculated. The advantage is the reciprocal of a polities attack power. Finally, the attack target is selected, where the probability of a polity being picked is equal to that polities advantage divided by the sum of the advantages of all of the neighbours.

**Figure 5** shows the imperial density at epochs 1500 BC–500 BC, 500 BC–500 AD and 500 AD–1500 AD using the greedy attack model. This attack model has a dramatic impact on the imperial density with large polities quickly spreading to all available cells at each epoch, even in areas that are expected to remain with relatively little development such as sub-Saharan Africa. $R^2$ values for the plots shown in **Figure 5** are given in **Table 1**, which confirm that this attack model less accurately models the historical prevalence and spread of large-scale polities.

The greedy attack model makes polities very effective at attacking and absorbing weak neighbours, so much so that the starting location of military technologies in the steppes no longer has a strong influence on the location of early large-scale polities or the path of the spread of such polities. To confirm this, the simulations using the greedy attack model were repeated with the starting locations of military technologies randomised. In these simulations, each cell has a 4.34% chance of beginning with all military technologies. This percentage is equal to the proportion of polity-supporting cells which are classed as steppes in the original simulation. With the starting locations of military technologies randomised, the simulations produce very similar results confirming that the influence of these starting locations on imperial density has been diminished.

When polities choose to attack randomly, the military technologies are the most important factor when building large scale polities as they increase the chance of ethnocide and hence the spreading of ultrasocietal traits which stabilises large polities. However, when polities attack weak neighbours preferentially, the attack success raises significantly making large scale polities more likely and less dependent on military technologies. This motivates the significant imperial density in areas far from the steppes (e.g., sub-Saharan Africa) andalso accounts for the greater imperial density on coastal cells when using the greedy model, as these cells have many more neighbours to exploit, or be exploited by via sea attacks. Indeed, if sea attacks are disabled this larger imperial density relative to landlocked cells disappears.

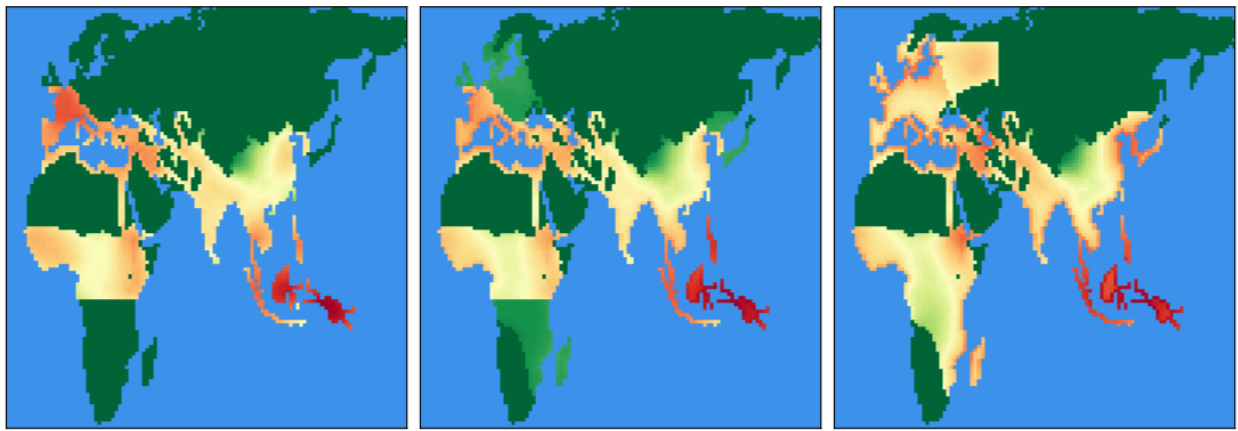

**Figure 5. Mean imperial density over 20 simulations using the greedy attack model at three epochs. From left to right 1500 BC–500 BC, 500 BC–500 AD and 500 AD–1500 AD.**

## Discussion and future work

In this report, we reproduce the results from Turchin et al. (2013), and release code and data for future use. Furthermore, we assess whether the model is predictive with respect to 1) population density as mapped by a city population dataset, finding a positive correlation; 2) areas of conflict as mapped by a dataset on historical battles, finding a mild correlation; 3) the shape of historical polities, finding no correlation. We further implemented a greedy version of the model, and verified the importance of random polity attacks, coupled with military technology diffusion from the steppes. To conclude, we discuss two aspects of future work: (1) proposed improvements and extensions to Turchin et al. (2013)'s original model; (2) a network science extension to the analysis of the results, to expand our understanding of complex societal dynamics.

Regarding the first point, we suggest two ways to improve the model, without changing its fundamental hypothesis. Currently, the conflict process forces each cell to attack each of its neighbours with equal probability. The model results are strongly dependent on this simplification, as we have shown. However, this is unrealistic when terrain, historical memory, and military factors are taken into account. Biasing the probability in accordance to a spatial interaction model (e.g., entropy maximization Boltzmann Lotka Volterra model in Wilson (2007)) could correct the spread of polities' shape towards more historically accurate directions. Furthermore, improvements to the underlying agricultural and terrain data resolution and accuracy from the Seshat database (Turchin et al. 2015) could also improve prediction accuracy.

Regarding the second point, creating a more formal understanding of the dynamic process via a spatial interaction network could yield further insights into the underlying spatial structure evolution of polities and resulting conflict. Connecting data-driven models with explicit generative models is important to enhance our understanding and predicting the posterior distribution of history-dependent processes and applications, e.g., peacekeeping (Guo et al. 2018). In our future work, we propose to create an evolving spatial interaction network model based on population data from Reba et al. (2016). By inferring the likelihood of interactions between populated settlements, we can create a weighted network. The community properties of the network across hierarchical spatial scales (e.g., strength of affiliation to a community, Karrer et al. (2011)) are likely to correspond to imperial density in Turchin et al. (2013). As such, we would expect the following: (a) robust network communities (e.g., cores, as in Seifi et al. 2013) to correspond to stable polities, and (b) unstable boundaries between communities to correspond to higher frequency of conflict (e.g., high betweenness, as in Guo et al. 2017). Probabilistically, the stochastic nature of imperial density and polity size (over Monte-Carlo simulation runs) in Turchin et al. (2013) maps well to the probabilistic community strength and sizes in most community detection algorithms.

## Acknowledgements and competing interests

The Alan Turing Institute authors acknowledge funding The Alan Turing Institute under the EPSRC grant EP/N510129/1 and the Defence & Securities Program under the Global Urban Analytics for Resilient Defence (GUARD) project. The authors have no competing interests.

## Contributions

J.M. conducted the simulation and result analysis. G.C., J.H., A.W. and W.G. oversaw the research process. J.M., G.C. and W.G. wrote the paper.. We also would like to thank James Bennett (J.B.) and Peter Turchin (P.T.) for their help in the source code explanations. Assistance and insight was provided by J.B., who had gained familiarity with the model in his own implementation, along with a listing of the original APL code from P.T.

## Code and data availability

The simulation code and data used to generate the results in this work is available on GitHub at https://github.com/alan-turing-institute/guard/releases/tag/v0.15. The notebooks directory of this repository contains Jupyter Notebooks which may be used to reproduce the figures presented here.

**Appendix**

## Implementation

We implemented the model from Turchin et al. (2013) from scratch in Python. This implementation of the model is written in an object-oriented style, which aids in creating clear definitions of entities in the simulation such as polities and individual cells. Abstract classes were also created for analysis of the simulation in the form of an accumulator class for collecting data over the course of a simulation (such as imperial density), and a correlator class for comparing an accumulator against external data (such historical population). The advantage of this design is that it greatly simplifies both making modifications to the simulation model and implementing new analyses. The trade off is that the code is not optimised to reduce simulation time and is likely considerably slower than a more traditional implementation using contiguous arrays of data. However, as individual simulations are short (2–3 minutes) we felt that the small burden of extra calculation time was more than compensated for by a clear and versatile code.

Input for the simulation software is provided in the form of YAML files. These are human readable text files, which again contributes to our approach of maximising clarity and usability. The creation and parsing of YAML files is also well facilitated in a range of programming languages through free and open source libraries.

In implementing the model a number of issues arose. First, two simulation parameters were changed from the values stated in the paper and supporting information, following Bennett (2016). $\varepsilon_{max}$, a parameter determining the probability of ethnocide occuring, was set to 2 compared to the value of 1 stated in the supporting information and $\Delta$, the amount that the sea attack distance is incremented each step, was set to 0.0025 instead of 0.025 as in the supporting information. All other parameters are equal to the stated values.

Second, the spread of military technology occurs in a manner different to that described in Turchin et al. (2013). Rather than each agricultural cell attempting to spread technology to a random neighbour each turn, the spread is actually very closely tied to attacks. In particular, attempts to spread military technology only occur after an attempt to attack, and always occur in the same direction of the attempted attack. The technology spread is attempted whether or not the attack was successful, or even if the attack didn't happen (*i.e.* when a cell attempts to attack another cell belonging to the same polity). Otherwise, the papers description of technology spread is consistent with the simulation, *i.e.* the random choice of a technology to attempt to share and the spread rate *a*.

Third, there are some peculiarities with sea attacks between littoral cells. To be consistent with the original simulation it is necessary that each littoral cell may attempt to attack itself when selecting a littoral neighbour to attack. These attacks are always rejected as cells are forbidden from attacking other cells belonging to the same polity. Furthermore, the distance between littoral cells, $d_{sea}$ in Turchin et al. (2013), is the euclidean distance between the centres of the two cells. The distance calculated does not necessarily correspond to a valid sea route between the two cells as it may cover land. Indeed, this means it is possible for any two littoral cells to attack each other even if there is no sea route between them (e.g. cells on the Caspian Sea attacking those in the Persian Gulf). However, as the sea attack distance remains small throughout the simulation (particularly in the light of the small increment described above) this is not considered to be a large source of unrealistic attacks.

Fourth, all cells must attack each of their neighbours with equal probability. Practically this is ensured by asserting that cells attack with equal probability in all four directions even if one or more of those directions is not a valid target for an attack such as a desert cell or a cell belonging to the same polity.

Finally, the order of events in each step was not described in the original work. The order in Turchin's simulation code, and our own, is:
1. Each cell (in a random order) attacks and attempts to spread military technology to its target.
2. Mutation of ultrasocietal trait vectors in each cell.
3. Polity disintegration.